\documentclass{epl}

\title{On a quantitative method to analyze dynamical and
measurement noise}
\shorttitle{Analysis of dynamical and
measurement noise}
\author{M.~Siefert\inst{1}
      \and A.~Kittel\inst{1}
      \and R.~Friedrich\inst{2}
      \and  J.~Peinke\inst{1}\thanks{%
      E-mail: \email{peinke@uni-oldenburg.de}}
}
\shortauthor{M.~Siefert \etal}
\institute{
    \inst{1} Carl von Ossietzky University, Physics Department -
             D--26111 Oldenburg, Germany\\
    \inst{2} Inst. Theor. Physik, University of
M\"unster , D--48149 M\"unster, Germany
}

\begin{document}

\maketitle

\begin{abstract}              

This letter reports on a new method 
of analysing experimentally gained time series with respect to 
different types of noise involved, namely,  we show that it is 
possible to
differentiate between dynamical and measurement noise. This method 
does not depend on previous knowledge of model equations. For the 
complicated case of a chaotic dynamics spoiled at the same time by 
dynamical and measurement noise, we even show how to extract from 
data the magnitude of both types of noise. As a further result, 
we present a new criterion to verify the correct embedding for chaotic dynamics with dynamical noise.

\end{abstract}
\section{\label{sec:Intro}Introduction}

Handling noise in experiments is a challenging task for an 
experimenter during everyday work regardless of the field he or 
she is working on.  Any knowledge of the nature of the involved noise is 
important to understand experimental results. It may help to estimate 
the achievable precision to make out noise induced effects or to set 
up models for the experimental system under investigation. For a 
general application it is essential that these methods should require 
as little knowledge as possible of the system.

In this 
paper we present evidence that it is possible for measured data, which 
were spoiled by different types of noise, to separate two basic types 
of noise and to measure their magnitudes. To show the quality of our 
method we apply it to the case of  a noisy nonlinear chaotic 
dynamical system. Obviously, this method also works for simpler 
dynamical situations, which are frequently given in experimental 
research.

Before the early 80s complex, disordered systems were explained predominantly 
by stochastic models.  The complex behavior of the dynamic was 
described by random motions. 
Then it became clear that many of these
disordered systems might be generated by low dimensional nonlinear
{\em deterministic} dynamics.  For both kinds of systems a lot of
refined methods for data analysis were
developed, cf. 
\cite{kantz,hilborn,silke,borland1,borland2,klimontovich,stark,muldoon}.
Complications in the data analysis based on this clear distinction arise
if noise is present beside a nonlinear deterministic dynamics. Two 
basic types of noise can be distinguished, namely, {\it dynamical 
noise}, which acts directly on the dynamics, and {\it measurement 
noise}, which is only added to the signal of the dynamics.
 Thus, 
for the analysis of disordered systems it is one fundamental
problem to characterize the type of noise and to quantify the amount of noise.

Recently, a method was proposed whereby dynamical noise and
measurement noise can be measured very precisely, if the dynamical 
equations are
known \cite{heald}. Our paper is devoted to the problem of unknown dynamics. It
is our intention to show that it is possible by pure data analysis to 
clarify which kind
of noise is present. Furthermore, by using the theory of diffusion processes one 
is
able to estimate the magnitude of dynamical and measurement noise. 
Our proceeding 
is based on recent works \cite{silke,christoph,malte,Lubijana} showing that it is possible to reconstruct from given data the underlying stochastic processes and we want to point out that it
 is not founded on any previous knowledge given by 
models
of the dynamic or by some assumed parameterizations.

The structure of the paper is as follows: First we describe the mathematics we
are using for the reconstruction of the deterministic flow in phase 
space from given data sets. Next it
follows the criterion for the distinction between measurement noise 
and dynamical
noise. We demonstrate that this method can be successfully applied to 
measured data of the chaotic Shinriki oscillator, which are perturbed 
by different types of noise. At last we show that the signal of the 
dynamical noise and its correlation can be extracted from the 
measured data. This can be taken to examine the nature of the 
stochastic process and to verify the sufficient high embedding of a 
chaotic noisy system.

\section{\label{sec:Concept} Concepts of stochastic processes}

Based on the mathematics  of diffusion processes it has recently been
realized that by directly  using the definition of the Kramers-Moyal 
coefficients
\cite{Kolmogorov,KramersMoyal} it is possible to reconstruct the dynamics
of the Langevin equation from given data
\cite{silke,christoph,malte,Lubijana}.
This idea is the foundation of the following presentation.

First we focus on the wide class of nonlinear dynamical
systems with dynamical noise, also known as the diffusion processes.
It can be represented by
a Langevin equation (in the It\^o representation),
\begin{equation}
\frac{d}{dt}X_i(t){=}{D_i^{(1)}({\bf
X},t)}+{\sum_{j=1}^n  \left[\sqrt{D^{(2)}({\bf
X},t)}\right]_{ij}\Gamma_j(t)},\;\;i=1,\ldots n
\label{langevin}
\end{equation}
where ${\bf X}(t)$ denotes the time dependent $n$-dimensional stochastic state
vector.  The drift coefficients, $D^{(1)}_i$, represent the
deterministic part of the dynamics, and the
diffusion coefficients, $D^{(2)}_{ij}$, determine the strength of the dynamical
noise, including the general case of multiplicative noise when the coefficients
$D^{(2)}_{ij}$ depend on ${\bf X}$.
$\Gamma_j(t)$ is $\delta$-correlated Gaussian noise (Langevin force).

   As known from \cite{Kolmogorov}, the drift coefficients $D^{(1)}_i$ 
are obtained
as the limit of conditional moments $M^{(1)}_i$
\begin{eqnarray}
D^{(1)}_{i}&=&\lim_{\Delta t \to 0} \frac{1}{\Delta
t}  M^{(1)}_{i}({\bf x},\Delta t) \\
M^{(1)}_{i}({\bf x},\Delta t)&=&\langle  X_{i}(t+\Delta t) - x_{i}(t)
\rangle\left.\right|_{{\bf X}(t)={\bf x}}\label{drift2}
\end{eqnarray}
and the diffusion coefficients  $D^{(2)}_{ij}$ by the moments $M^{(2)}_{ij}$
\begin{eqnarray} 
D^{(2)}_{ij}&=&\lim_{\Delta t \to 0} 
\frac{1}{\Delta t}M^{(2)}_{ij}({\bf x},\Delta t) 
\label{diffusion}\\
M^{(2)}_{ij}({\bf x},\Delta t)&=&\langle\big( X_{i}(t+\Delta t) -
x_{i}(t)\big)\big( X_{j}(t+\Delta t) - x_{j}(t)\big) 
\rangle\left.\right|_{{\bf X}(t)={\bf x}}.
\label{diffusion2}
\end{eqnarray}
The numerical estimations of these conditional moments are
performed for  ${\bf X}(t)\in U({\bf x})$, for a sufficiently small
neighborhood $U$ of a fixed value ${\bf x}$ in the phase space. 
These conditional
moments can be estimated directly from given data sets in a parameter
free way.  For small
$\Delta t$ (i.e. smaller than the recurrent time \cite{TimmerdelT}) the first
two moments $M^{(i)}$ ($i=1,\,2$) are connected to the
diffusion coefficient \cite{risken,commentM2}:
\begin{eqnarray}
     M_{ij}^{(2)}({\bf x},\Delta t)
     -M_{i}^{(1)}({\bf x}, \Delta t)M_{j}^{(1)}({\bf
     x}, \Delta t)=D_{ij}^{(2)}({\bf x})\Delta t + O(\Delta t^2).
\end{eqnarray}

If, in addition to the dynamical noise, also 
measurement noise is present the procedure of the estimation of 
$D_{ij}^{(2)}({\bf x})$ has to be changed.
The measurement noise, 
which is typically added by the process of
measuring data, can be formulated mathematically as
\begin{equation}
	Y_i(t)=X_i(t)+\sigma_{i} \zeta_i(t). \label{measurement}
\end{equation}
The vector $Y_i$ is the sum of the state vector $X_i$ described by the
dynamics of Eq. (\ref{langevin}) and measurement noise. Here the 
measurement noise is given by its standard deviation $\sigma_{i}$ and 
the
$\delta$-correlated noise term $\zeta_i$. As a consequence of the 
definition (\ref{measurement}), it is easy to see that for ${\bf y}$ the conditional moments, as defined in Eqs. (\ref{drift2}) and (\ref{diffusion2}), one obtains
\begin{eqnarray}
     K_{ij}^{(2)}({\bf y},\Delta t)&:=&M_{ij}^{(2)}({\bf y},\Delta t)
     -M_{i}^{(1)}({\bf y}, \Delta t)M_{j}^{(1)}({\bf
     y}, \Delta t)\nonumber \\&=&
D_{ij}^{(2)}({\bf x})\Delta t+2\sigma^2_{i}\delta_{ij} +
				O(\Delta t^2).\label{k2}
\end{eqnarray}
Note that for the determination of $D_{i}^{(1)}({\bf 
x})$ via $M_{i}^{(1)}({\bf y})$ (see equation (\ref{drift2})) no 
correction term appears due to the measurement noise, because it 
averages out.

The equation (\ref{k2}) is valid for a sufficient small neighborhood 
$U({\bf x})$ so that
$M_{i}^{(1)}$ and $M_{ij}^{(2)}$ can be approximated by constant 
values in $U({\bf x})$.  Furthermore the linear dependence of 
$K^{(2)}$ on $\Delta t$ can be taken as a criterion for  a correct 
sampling frequency, which has to be chosen so high that this 
linearity is resolved.

\begin{figure}
\onefigure[width=3in]{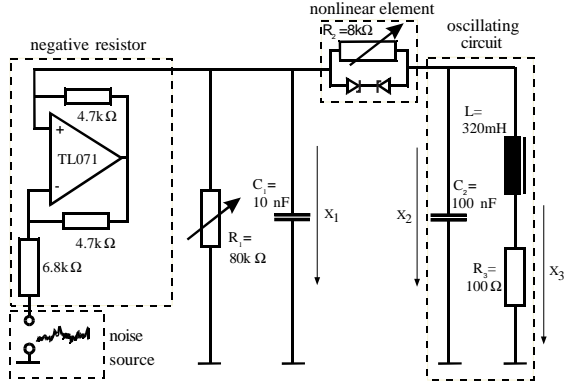}
\caption{Circuitry of the Shinriki oscillator. The noise source placed in series
   to the negative resistor.}
\label{fig.circuitry}
\end{figure}
\begin{figure}[]
   \begin{center}
\includegraphics[width=2in]{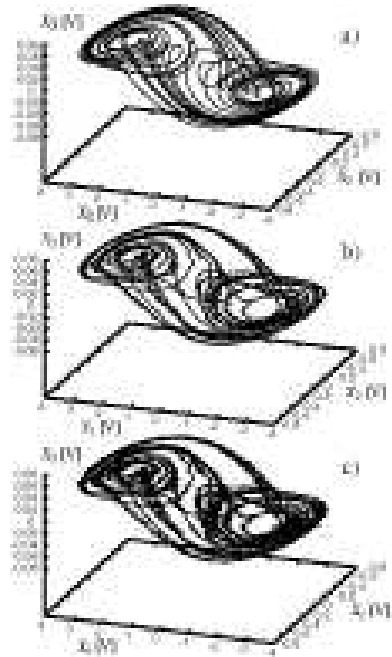}
   \end{center}
   \caption{Trajectory for the Shinriki oscillator in the phase space
   with different kind of perturbing noise.  a) without noise, b) with
   dynamical noise ($[\sqrt{D^{(2)}}]_{11}$=7.9 V/$\sqrt{s}$), c) with
   dynamical noise (like in part b)) and measurement noise 
($\sigma$=0.12 V).\label{fig.attractor}
     }
\end{figure}

Next we apply the method to measured data of a chaotic
electronic oscillator.  As a circuitry we have chosen the Shinriki
oscillator \cite{shinriki} as shown in Fig.  \ref{fig.circuitry}.
In Fig.  \ref{fig.attractor} exemplary
phase space representations of the attractors for the measured data are
shown.  Fig.
\ref{fig.attractor}a) shows the pure deterministic chaotic dynamics, Fig.
\ref{fig.attractor}b) the dynamics  with dynamical noise, and
  Fig. \ref{fig.attractor}c) dynamics with the combination of 
dynamical and measurement noise.
For an experimental realisation of the dynamical noise, a $\delta$-correlated noise source is in series connection to the
negativ resistor.
The  corresponding Langevin equation for the three voltages $X_i$, 
describing the Shinriki
oscillator, see Fig. \ref{fig.circuitry}, are given by
\begin{eqnarray}
\dot X_1 &=& - \frac{X_1 - \Gamma_1(t)}{R_{N}C_{1}}-
\frac{X_1}{R_1C_{1}}-\frac{f(X_1-X_2)}{C_1}
\label{shin1}\\ &=& g_{1}(X_{1},X_{2})+h_{1}
\Gamma(t)  \nonumber\\
\dot X_2&=&
\frac{f(X_1-X_2)}{C_2}-\frac{1}{R_3C_{2}}X_3 = 
g_{2}(X_{1},X_{2},X_{3}) \label{shin2}
\\
\dot X_3&=&-\frac{R_3}{L}(X_2-X_3) = g_{3}(X_{2},X_{3}),
\label{shin3}
\end{eqnarray}
where $h_{1}\Gamma(t)$ describes the Langevin force. 
For the specific parameters see Fig. \ref{fig.circuitry}, the 
negative resistor $R_N =-6.8k \Omega $ and $f(\cdot)$ describes the 
nonlinearity of the Zener diodes.
 An empirical formular for the characteristic curve is
\begin{equation}
	f(V)=\left\{
	\begin{array}{cc}
	\textrm{sign}(V)(A(\Delta V)(\Delta V)^{2}
	+B(\Delta V)^{3}+C(\Delta V)^{5})&\textrm{if}\;\Delta V>0\\
	0&\textrm{else}
	\end{array} \right.\label{fd}
\end{equation}
where $\Delta V=|V|-V_{D}$. The four parameters $A$, $B$, $C$ and $V_{D}$ have to be fitted on the measured characteristic curve.
Additional measurement noise was 
added to the data, namely to the component $X_1$.

\begin{figure}[]
   \begin{center}
\includegraphics[width=2in]{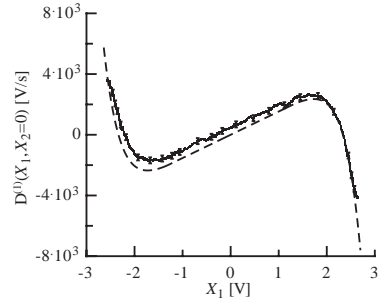}
   \end{center}
   \caption{The experimentally estimated drift coefficient 
$D^{(1)}_{1}(X_{1},X_{2}=0)$ of 
the Shinriki
   dynamic with error bars. Broken curve - measurement of the 
corresponding circuit elements.
 \label{fig.d1shinriki}
     }
\end{figure}

To give evidence of the validity of our procedure for the case of dynamical
noise, we show in Fig. \ref{fig.d1shinriki} the reconstructed deterministic
part of Eq. (\ref{shin1}), which we obtained from measured data (here and in the following we use 400.000 data points for our analysis).
Here an exemplary cut
through $\{ \bf D^{(1)}, \bf X \} $ has been chosen in
such a way that the nonlinearity becomes obvious.  By measuring the 
electronic elements ($R_N, C_1, R_1$ and $f(\cdot)$) we can directly 
compare the characteristics gained from Eq.  (\ref{shin1}) with the 
reconstructed one (see Fig.  \ref{fig.d1shinriki}). The small 
deviations can be explained by parasitic capacitances and inductances. For analogous numerically generated data sets no significant deviation of the reconstructed values of $D^{(1)}$ was found.

\begin{figure}[]
   \begin{center}
   \includegraphics[width=2in]{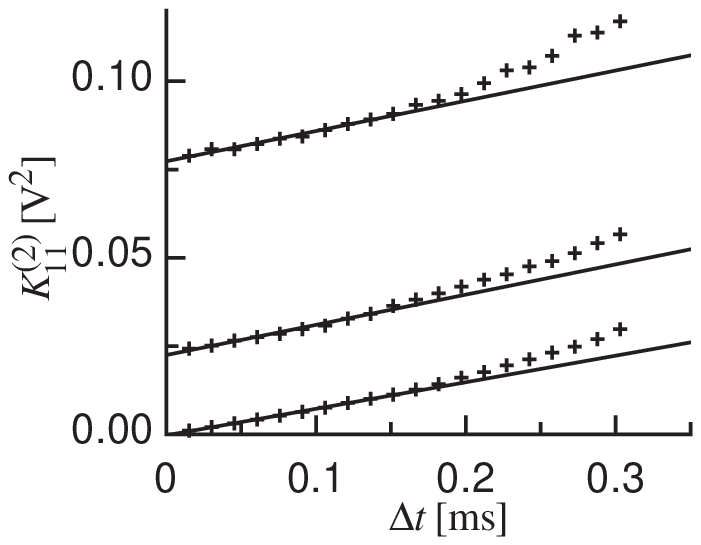}
   \end{center}
   \caption{ \label{fig4}
   The second cumulant $K^{(2)}(\Delta t)$ of Eq. (\ref{k2}) for the 
Shinriki oscillator perturbed by dynamical and measurement noise. The 
different sets of data are obtained for
   increasing amplitudes of measurement noise (bottom up: $\sigma=$0, 0.11, 0.2 V). The 
straight lines show linear fits, from which the slope ($D^{(2)}$) and 
the offset ($2\sigma ^2$) are obtained.
\label{fig.limesshinriki}
     }
\end{figure}

Furthermore, we investigate the diffusion coefficients.
For simplification, only
  the case of additive noise is  considered, i.e. $D^{(2)}$ is constant.
According to equations (\ref{drift2}), (\ref{diffusion2}) and (\ref{k2}) we
calculate
$K^{(2)}(\Delta t)$.  To improve the statistics
we calculate the median of $K^{(2)}(\Delta t)$ about the whole state space.
As shown in Fig. \ref{fig.limesshinriki} the moments $K^{(2)}$ display a
linear dependence on small $\Delta t$ \cite{comment}.  The slope of 
this dependence gives
  the strength of the dynamical noise $D^{(2)}$, see Eq. (\ref{k2}).
Most remarkably $K^{(2)}$ shows an increasing off-set when the 
measurement noise is increased.  According to
equation (\ref{k2})  with the value of $K^{(2)}(x,\Delta t =0)$
  the strength of the measurement  noise $\sigma_{i}$ can be 
measured. Our results are summarized in
table \ref{tab.shinriki}. The precision of these results obviously depends 
on the number
of data points. Furthermore we notice that with
increasing magnitude of the measurement noise the value $\sigma$ gets
underestimated while the precision of the estimated $D^{(2)}$ almost remains about
constant. ($\sigma = 0.24$ corresponds to about $4 \%$ noise.)

\begin{table}
\caption{ Values of the measurement noise 
$\sigma$ and the diffusion coefficient $D^{(2)}_{11}$ as adjusted in 
the experiment and estimated from the measured 
data.}
\label{tab.shinriki}
\begin{center}
\begin{tabular}{cccc}
$\sigma$ [V] adjusted  &0.0&0.12&0.24\\
$\sigma$ [V] estimated &-0.011$\pm 0.01$&0.11$\pm 0.01$&0.20$\pm 0.01$\\
${\sqrt{D^{(2)}}}_{11}$ [V/$\sqrt{\hbox{s}}$] adjusted &7.9&7.9&7.9 \\
${\sqrt{D^{(2)}}}_{11}$ [V/$\sqrt{\hbox{s}}$] estimated &8.4
$\pm 0.3$&8.5$\pm 0.3$&8.4$\pm 0.3$\\
\end{tabular}
\end{center}
\end{table}

An  important consequence of this method should be noted.  In the case of
pure dynamical noise it is easy to see from Eq. (\ref{langevin}) that 
the knowledge of
$D^{(1)}$ and $D^{(2)}$ makes it possible to extract from measured 
data the noise term $\Gamma(t)$.
Based on this, it can be quantified 
whether the noise
is $\delta$-correlated or not. As an illustration the autocorrelation
of the reconstructed noise  is
shown in Fig. 5a). Note that correlations are expected if the 
inserted noise is not
$\delta$-correlated. To investigate such a case we use a too low 
dimensional phase space embedding of our measured data. In Fig. 5b) 
the
autocorrelation of the reconstructed noise is shown for the case that 
the data of Fig
5a) are reduced to a two-dimensional projection  of the dynamics on 
$X_1$ and $X_3$.  In this case the unresolved variable $X_2$ together 
with $\Gamma_1$ represent  correlated noise.
This result clearly 
shows two points: (a)  the validity of a Markov process (i.e. the noise is 
$\delta$-correlated) can be 
verified; (b) if correlations are found, like those shown in figure 
5b), the system does not obey a Markov process.

\begin{figure}[]
   \begin{center}
\includegraphics[width=2in]{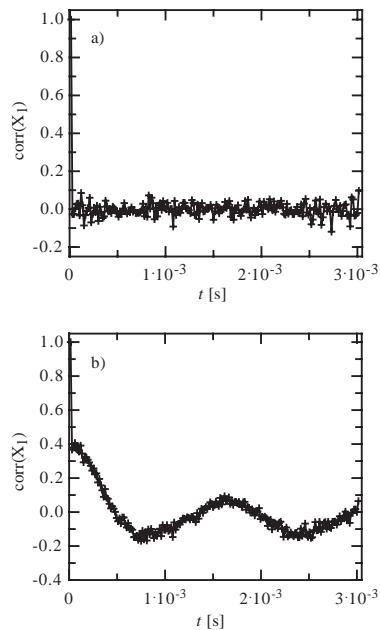}
   \end{center}
   \caption{ \label{fig5}
Autocorrelation function of reconstructed dynamical noise, a) 
correctly embedded showing
$\delta$-correlated noise, b) the projected dynamics in the two-dimensional phase space
${X_1}(t)$ and ${X_3}(t)$, showing finite time correlations.}
\end{figure}

To conclude, in this paper we show for the first time, that
based on the well known theory of diffusion
processes, and especially based on the estimation of the Kramers-Moyal
coefficients it is possible to analyze the kind of noise given in time series.  The method shown here does not depend on
previous knowledge of the underlying nonlinear deterministic dynamics.
This does not imply that our method must work for any dynamical 
process. From an experimental point of view, the obtained results 
have to be verified whether the correct dynamics is grasped by the 
reconstructed process. Therefore the acting noise can be 
extracted and the typical dynamics can be obtained by numerical 
integration of the reconstructed phase flow using the obtained values 
of $D^{(1)}$. If this is successful, a further improvement of the 
estimation of the reconstructed process can be achieved by 
parameterizing the results of our method and successively applying 
procedures for parameter estimation like \cite{heald,TimmerCSF}.


\acknowledgments
Helpful discussions with S. Siegert, Ch. Renner, H. Kantz  
as well as the hospitality of the Max-Planck Institute for the 
Physics of Complex
Systems, Dresden are acknowledged.


\begin{thebibliography}{0}
\bibitem{TimmerdelT} Here we used 77.5 samples per mean recurrent time,
corresponding to a sampling rate of 66 kHz. Problems arising for large $ \Delta
t$ are discussed in \cite{TimmerCSF}.
\bibitem{commentM2} An expansion of $M^{(2)}$ to 
higher order terms in $\Delta t$ is given in 
\cite{Kantz2,Com}.
\end{thebibliography}
\end{document}